\documentclass[%
reprint,
superscriptaddress,
amsmath,amssymb,
aps,
]{revtex4-2}

\usepackage{graphicx}% Include figure files
\usepackage{dcolumn}% Align table columns on decimal point
\usepackage{bm}% bold math
\usepackage{xr}
\usepackage{color}

\definecolor{waldgruen}{RGB}{34,139,34}

\begin{document}
	
%	\preprint{APS/123-QED}
	
	\title{Mean-field models for the chemical fueling of transient soft matter states}
    
	%\thanks{A footnote to the article title}%
	
	\author{Sven Pattloch}
	\affiliation{Applied Theoretical Physics - Computational Physics, Physikalisches Institut, Albert-Ludwigs-Universit\"at Freiburg, D-79104 Freiburg, Germany}
	\affiliation{Cluster of Excellence livMatS @ FIT - Freiburg Center for Interactive Materials and Bioinspired Technologies, Albert-Ludwigs-Universit\"at Freiburg, D-79110 Freiburg, Germany}
   	\author{Joachim Dzubiella}
	\affiliation{Applied Theoretical Physics - Computational Physics, Physikalisches Institut, Albert-Ludwigs-Universit\"at Freiburg, D-79104 Freiburg, Germany}
    \affiliation{Cluster of Excellence livMatS @ FIT - Freiburg Center for Interactive Materials and Bioinspired Technologies, Albert-Ludwigs-Universit\"at Freiburg, D-79110 Freiburg, Germany}
	
	\date{\today}% It is always \today, today,
	%  but any date may be explicitly specified
	
\begin{abstract}
The chemical fueling of transient states (CFTS) is a powerful process to control the nonequilibrium  structuring and the homeostatic function of adaptive soft matter systems. Here, we introduce a mean-field model of CFTS based on the activation of metastable equilibrium states in a tilted `Landau' bistable energy landscape along a coarse-grained reaction coordinate (or `order parameter') triggered by a nonmonotonic two-step chemical fueling reaction. Evaluation of the model in the quasi-static (QS) limit - valid for fast system relaxation - allows us to extract useful analytical laws for the critical activation concentration and duration of the transient states in dependence of physical parameters, such as rate constants, fuel concentrations, and the system's distance to its equilibrium transition point. We apply our model in the QS limit  to recent experiments of CFTS of collapsing responsive microgels and find a very good performance with only a few global and physically interpretable fitting parameters, which can be employed for programmable material design. Moreover, our model framework also allows a thermodynamic analysis of the energy and performed work in the system. Finally, we go beyond the QS limit, where the system's response is slow and retarded versus the chemical reaction, using an overdamped Smoluchowski approach. The latter demonstrates how internal system time scales can be used to tune the time-dependent behavior and programmed delay of the transient states in full nonequilibrium.  
\end{abstract}
	
%\keywords{Suggested keywords}%Use showkeys class option if keyword
%display desired
\maketitle
	
%\tableofcontents

% -------------------------------------------------------
% -------------------------------------------------------
\section{Introduction}

The transient assembly and ordering of active materials fueled by a chemical reaction is a key process in the nonequilibrium structuring and function of biomolecular systems, e.g., to perform work or reach homeostatic mechanical responses~\cite{desai,whitesides}. These versatile and adaptive material features have triggered plenty of research recently, on one hand, to understand the fundamental physical properties of nonequilibrium transient states, but also, on the other hand, to develop synthetic active materials which display biomimetic or other novel useful behavior, driven by fuel consumption through chemical reaction networks~\cite{Balazs,Siegel, Merindol_2017,Walther_Roadmap, Wang_2021, Hoefling, Piazza, Sharko_2022}. Experimental examples are the fuel-driven self-assembly of synthetic molecules into fibers~\cite{Boekhoven_2015} or gels~\cite{heuser,Panja_2019} with variable and controllable lifetime and stiffness, the fueled nucleation and coacervation~\cite{deng} and spinodal decomposition~\cite{heckel} in phase separating systems, as well as the fueled collapse of functional macromolecules such as hydrogel colloids~\cite{Heckel_2021,Nakamoto_2022}.

The desired goal of the ongoing research efforts is to establish rational design principles that enable a generic access to nonequilibrium soft matter systems with adaptive and predicable dynamics~\cite{Walther_Roadmap,Heinen_2019}, for example,  to demonstrate programmable hydrogel-based model systems~\cite{Zhang_2019,Klemm_2022}. Hydrogels are soft, responsive and deformable, and thus of special interest for the development, e.g., of  chemically fueled mechanical actuators \cite{Ionov_2014, Fusi_2023}.  However, realizing programmable or even adaptive structural dynamics has proven challenging because it requires harmonization of the chemical energy uptake and dissipation events within the steady states~\cite{Piazza}. The  full nonequlibrium is even more difficult to control due to the intricate coupling of the time-dependent chemical, thermodynamic, as well as mechanical  degrees of freedom of the supramolecular systems~\cite{Balazs,Siegel, Aizenberg}.  The theoretical modeling is therefore often either too complex to derive simple laws, or relies only on the numerical solution and phenomenological interpretation of the underlying chemical networks~\cite{Postma_2017, heckel, Nakamoto_2022, Sharko_2022} without coupling to low-dimensional emerging mechanical or structural (order) parameters during the spatiotemporal evolution of the whole system. 

Here, we make a first step towards a simple theoretical treatment of the coupling of the chemical fueling to the emerging structure, thermodynamics and mechanics of the system within a generic model framework. The latter is motivated by a Landau-type of mean-field model to access the qualitative behavior of phase transitions, e.g., as of magnetic systems in external fields~\cite{Landau}. In particular, we assume that the fueled system is bistable (two-state), featuring a stable state and a highly unstable state, the latter of which is then activated by the external field. In other words, fueling increases the probability of the unlikely `hidden' state over the initial state for a certain time, thus stabilizing a transient state with variable lifetime.  In contrast to the classical Landau model, the external field enters in our approach through the action of a time-dependent chemical reaction (or chemical network). As a first approximation, the field enters linearly into our model analogous to the popular $m-$value approach to describe biomolecular state transitions, such as 2-state protein unfolding/denaturation by cosolute addition~\cite{Pace_1975,Schellman_1978,Heyda_2014}. Although simple and mean-field, we demonstrate that many useful scaling laws can be drawn from such a model already in the quasi-static limit (where system relaxation is fast compared to the chemical reaction), in particular, for the relations between fuel concentration, chemical rates, and the duration of the transient states. Moreover, we show that, like in the Landau framework~\cite{Landau}, such a Hamiltonian-based model can then also be employed by using simple diffusive relaxational dynamics to study a full nonequilibrium fueling process. This is relevant in situations when the chemical and the system time scales are comparable and temporal effects like delay and retarded response come into play. We discuss further possible applications and extensions of these models in the final outlook section of this work. 

% -------------------------------------------------------
% -------------------------------------------------------
\section{General model}
%Formulas

\subsection{Coarse-grained bistable Hamiltonian}
In our model, the fueled system is described by a coarse-grained one-dimensional reaction coordinate, $Q$, e.g., the radius of a single responsive particle, cf. Fig.~1, or, in general, any meaningful structural (order) parameter.  In order to allow for state transitions of the system (as, e.g., in the hydrogel volume transition~\cite{Emanuela, Fernandez-Rodriguez_2023}), the coordinate is assumed to live in a bimodal energy landscape, ${\cal H}_0(Q)=A(\Delta Q)^2+B(\Delta Q)^4$, which we model by a simple quartic form as put forward in the simplest case by Landau to model phase transitions~\cite{Landau}. Here, $\Delta Q = Q-Q_c$, and $A$, $B$, and $Q_c$ describe the intrinsic energy landscape, with $Q_c$ being the center of the symmetric quartic form.  For $A<0$ and $B>0$ it exhibits two local minima at $Q_1$ and $Q_2$. If $Q$ is, for example, a particle volume or size, then the interpretation of such an Hamiltonian would be that it essentially represents a nonlinear elastic energy including a volume transition.  

The action of the chemical fuel is considered by a time-dependent contribution ${\cal H}_p(Q,t)=m \cdot \left(p(t)-p^* \right)\Delta Q$, which constitutes a perturbation of ${\cal H}_0$ linear in both $Q$ and in the product concentration $p(t)$, like an external magnetic field in the Landau picture. The total form of the Hamiltonian is thus
\begin{eqnarray}
\label{eq:Hm}
	{\cal H}(Q,t)&=&{\cal H}_0(\Delta Q)+{\cal H}_p(\Delta Q,t) \\
	& =& A(\Delta Q)^2+B(\Delta Q)^4+ m \cdot \left(p(t)-p^* \right)\Delta Q. \nonumber
\end{eqnarray}
 The value of $m$ defines the strength of the action of the chemical products $p$. For the critical concentration $p^*$ the chemical contribution ${\cal H}_p(Q)$ vanishes and the two coarse-grained states are equally probable. In other words, $p^*$ describes the initial bias (tilt) of the bimodal landscape in the unperturbed equilibrium.  The Hamiltonian (\ref{eq:Hm}) is explicitly time-dependent because of the time-dependent product concentration $p(t)$. The increase of the latter leads to large tilts of the landscape, activating metastable states into transient probable states, cf. Fig.~1.  

We note that a linear perturbation of thermodynamic two-state energies is not always justified but it is simple and can almost always be derived from a Taylor expansion as weak perturbation. It is thus quite established within the `$m$-value' framework for the action of simple cosolutes on the coil-to-globule (or folding/unfolding) transition in bio- and polymer physics~\cite{Pace_1975,Schellman_1978,Heyda_2014}). In that sense $mp^*$ can also be interpreted as a thermodynamic distance to the (coil-to-globule or volume) transition temperature $T_{\rm crit}$ and is essentially related to a temperature difference $\propto T-T_{\rm crit}$ times transition entropy~\cite{Heyda_2014}. In other words, for a certain responsive system with a thermodynamic transition temperature, the initial tilt $p^*$ can be pre-designed by temperature, transition entropy, and the chemistry specific $m$-value, which are often known or measurable quantities.

\begin{figure}[h]
	\centering
		\includegraphics[width=0.45\textwidth]{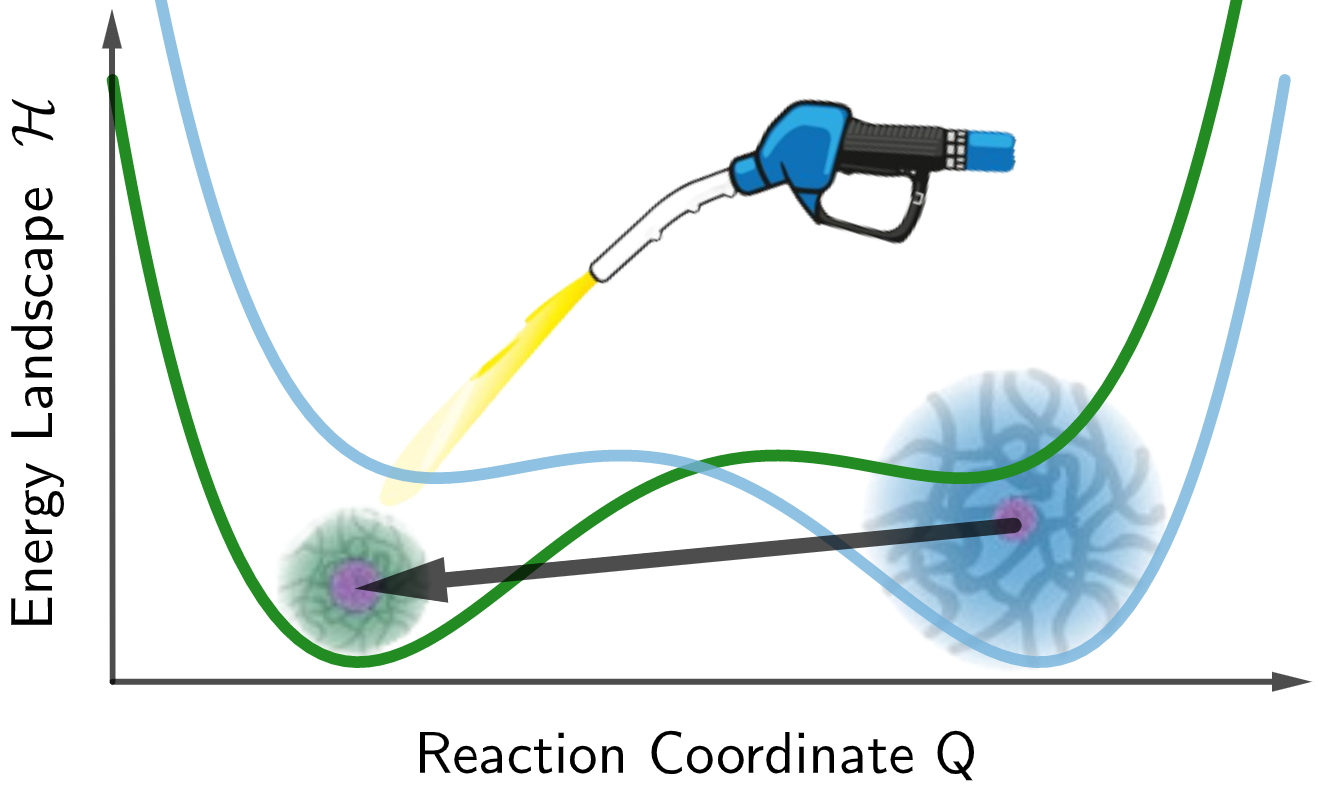}
	\caption{Landau like energy landscape ${\cal H}(Q)$ along a reaction coordinate $Q$, exemplified for the radius $R$ of a spherical hydrogel particle with small (greenish hydrogel) and big (blueish hydrogel) states. The fueling leads to a tilt of the equilibrium landscape (blue line) and transiently stabilizes the small state in the activated landscape (green line).}
	\label{fig:Scheme}
\end{figure}

\subsection{Chemical fueling reaction}
The chemical fueling is assumed to follow a two-step reaction process. Here, the fuel, $f(t)$, is converted in the homogeneous solution to a product, $p(t)$, with a rate constant $\tilde k_+$, following the rate equations 
\begin{equation}
\label{eq:c5}
\begin{split}
	\dot{f} &= -\tilde k_{+} f \left(p_\mathrm{sat} - p \right)\\
	\dot{p} &= \phantom{-}\tilde k_{+} f \left(p_\mathrm{sat} - p \right) - k_{-}p
\end{split}
\end{equation}
with the starting conditions $f\left(t=0 \right) = f_0$ (i.e., the initial fuel concentration) and no product initially, $p\left(t=0 \right)= p_0 = 0$. The product is the species that is active, in the sense that it changes the system by interacting or physically/chemically binding to it. Typically, the product has only a certain lifetime and decays with a first order rate $k_-$. Note that the $\tilde k_+$ is a second order rate, which we denote by the tilde symbol. Moreover, we have to impose a saturation concentration, $p_{\rm sat}$ for the action of the fuel to consider the possibility of a finite number of products because of, e.g., limited binding partners/sites. In equilibrium, $\dot p=0$, we recover the Langmuir isotherm for the function $p(f)$ with equilibrium constant $\tilde k_+/k_-$~\cite{langmuir}.  During time evolution, the products reach a single maximum, $p_{\rm max} := \max(p) \leq p_{\rm sat}$, as further exemplified below. 

{Generally, we can distinguish various regimes depending on whether the ratio of fuel to $p_{\rm sat}$ and the ratio of $\tilde k_+ p_{\rm sat}$ to $k_-$ is low/high. In a related discussion by Sharko {\it et al.}~\cite{Sharko_2022} it is shown that to activate the transient state, we need either very high fuel concentrations, or higher activation than deactivation rates in order to obtain sufficiently many active products. We will quantify this more in the following for our model. Since we focus on systems with controllable lifetime we restrict ourselves to the activation dominant case $\tilde k_+ p_{\rm sat} > k_-$.}
%We naturally restrict the rates that always $\tilde k_+ p_{\rm sat} > k_-$, i.e., fueling is faster than the following decay. 

\begin{figure}[h]
	\centering
		\includegraphics[width=0.45\textwidth]{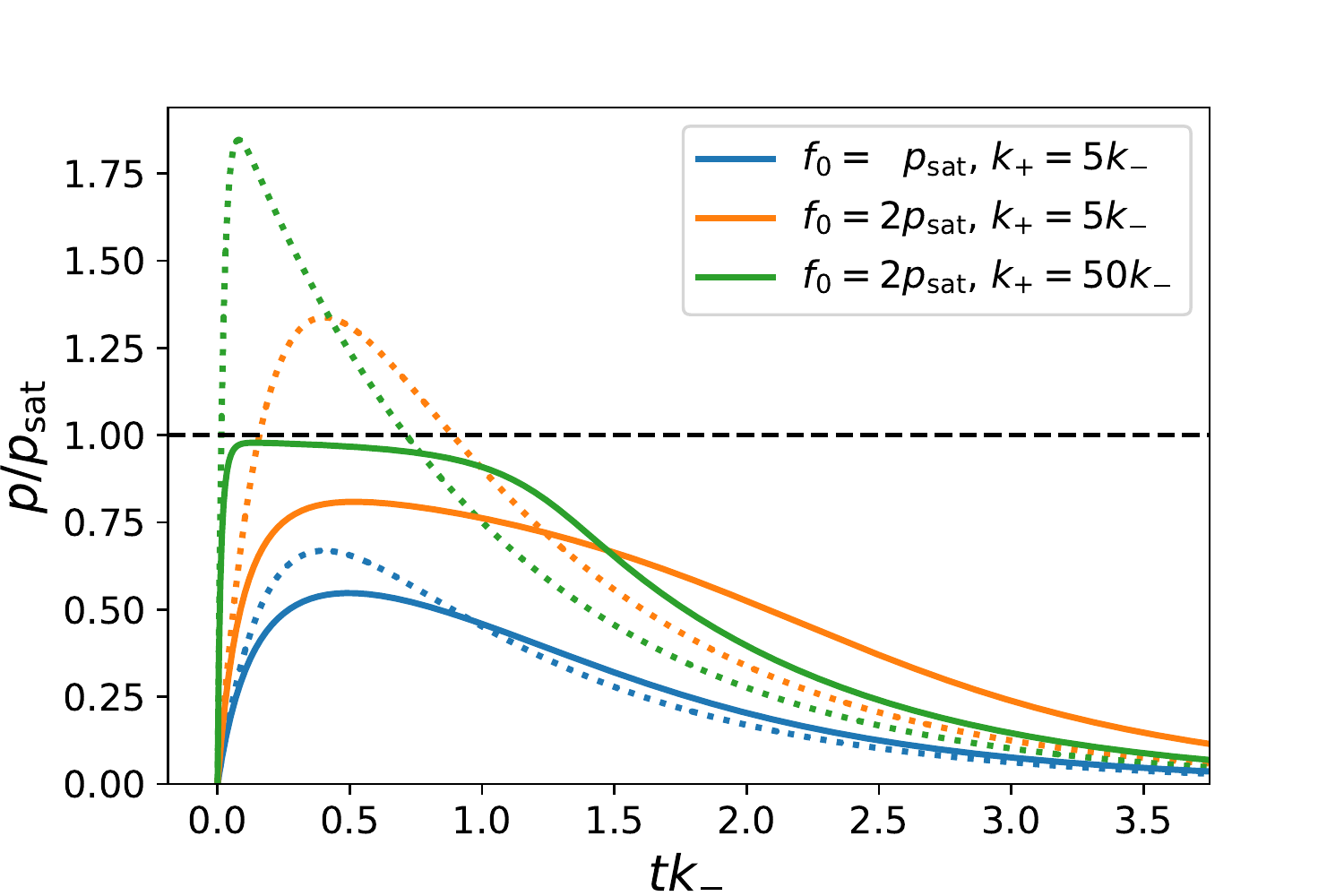}
	\caption{Product concentration $p(t)$ (scaled by $p_{\rm sat}$) for different parameter settings.  Solid lines are according to the numerical solutions of eq.~(\ref{eq:c5}). Dotted lines show the concentration profiles for the same parameters without saturation, following the analytical solution, eq.~(3). 
		Besides the initial fuel concentration $f_0$ we vary $k_{+}$, cf. legend. }
	
	%The green and the orange line neglect the saturation term, while the other ones take it into account. Besides the initial fuel concentration $f_0$ we vary the remaining parameters $k_{+}$ and $k_{-}$ in the purple and brown curves, cf. legend.}
	\label{fig:c_comp}
\end{figure}

Analytic solutions for $p\left(t \right)$ we only obtain for the unsaturated case $p_\mathrm{sat}\gg p$ for all times, where we can approximate $p_{\rm sat} -p \simeq p_{\rm sat}$ and introduce a new pseudo first order rate constant $k_+ = \tilde k_+p_{\rm sat}$. The analytical solution is then analogous to the double-exponential two chain reaction of radioactive decay~\cite{Bateman_1910}
\begin{equation}
\label{eq:c2}
\begin{split}
	p\left(t \right) %&=  \frac{ k_{+}f_{0}}{ k_{+}-k_{-}} e^{-k_{-}t}-\frac{ k_{+}f_{0}e^{- k_{+}t}}{ k_{+} -k_{-}}\\
	&= \frac{ k_{+}f_{0}}{ k_{+}-k_{-}} \left(e^{-k_{-}t} - e^{- k_{+}t}\right).
\end{split}
\end{equation}
In this case, the time evolution of products $p(t)$ shows an exponential rise with rate $ k_+$ at the beginning, a maximum at $p_{\rm max}$ at $t=t_{\rm max}$, following a decay with rate $k_-$ for long times. This is exemplified in Fig.~\ref{fig:c_comp}, where we compare different fueling situations.  The time where the product concentration is maximal in the unsaturated (us) case is given by 
\begin{equation}
\label{eq:cmax}
\begin{split}
	t_\mathrm{max}^{(\rm us)} &= \frac{\ln \kappa}{k_{-}- k_{+}}
\end{split}
\end{equation}
with the corresponding maximum product concentration  
\begin{equation}
\label{eq:ctmax}
\begin{split}
	p_{\rm max}^{(\rm us)} = p(t_{\rm max}^{(\rm us)}) = \frac{f_{0}}{1-\kappa} \left(\kappa^{\frac{\kappa}{1-\kappa}} - \kappa^{\frac{1}{1-\kappa}}\right),%, \left(e^{\frac{\kappa\ln{\kappa}}{1-\kappa}} - e^{\frac{\ln{\kappa}}{1-\kappa}}\right)
\end{split}
\end{equation}
where we introduced 
$\kappa = {k_{-}}/{ k_{+}}$, the ratio of the rates with $0<\kappa<1$. The analytical solutions for the unsaturated case are compared to numerical solution of eq.~(2) for saturated situations in Fig.~2. The saturated cases in these examples show suppressed peaks and plateau-like behaviors where $p\leq p_{\rm max} < p_{\rm sat}$ and always $ p_{\rm max} < p_{\rm max}^{(\rm us)}$. 

\subsection{Quasi-static (QS) chemical fueling}
\subsubsection{Equilibrium averages}
If the reaction coordinate $Q$ relaxes much faster than the chemical timescales, the dynamics are quasi-static (QS), i.e., the Boltzmann distribution of $Q$ 
\begin{equation}
P(Q,t) = \exp(-\beta {\cal H}(Q,t))/Z(t)
\end{equation}
according to the Hamiltonian (\ref{eq:Hm}) holds for every time $t$.  The normalizing partition sum
is \begin{equation} 
Z(t) = \int e^{-\beta {\cal H}\left(Q, t\right)} \mathrm{d}Q.
\end{equation} 
The average value of a function $X(Q,t)$ then directly follows from the Boltzmann average
\begin{equation}
\label{eq:P}
	\langle X(t)\rangle = \int X(Q,t)\,P(Q,t)\, \mathrm{d}Q.
\end{equation}
Hence, in the QS limit we can straightforwardly consider also thermodynamic quantities such as energy $U(t) = \langle {\cal H}\rangle$, free energy $F(t)=-k_\mathrm{B}\ln Z(t))$,  entropy $S(t) =(U(t)+F(t))/T$, and the power $P=\mathrm{d}F/\mathrm{d}t$. Using the exact relation $\mathrm{d}F/\mathrm{d}t = \langle \mathrm{d}{\cal H}/\mathrm{d}t\rangle$, one can derive the useful relation for our model that the power is 
\begin{equation}
\label{eq:power}
P = m (\mathrm{d}p/\mathrm{d}t) \langle \Delta Q(t)\rangle,
\end{equation}
 i.e., given by the change of time evolution of the product times the time-dependent mean of the order parameter. The initial fueling power is then provided by $P(t=0) \simeq m f_0 k_+ (Q_2-Q_\mathrm{c})$ if we use $\langle \Delta Q(t=0)\rangle \simeq Q_2-Q_\mathrm{c}$. 

\subsubsection{Separation of the energy contributions}
The time-dependent work and energy can be deeper analyzed by considering the contributions to the Hamiltonian, eq.~(\ref{eq:Hm}). It consists of two parts. First, we have the intrinsic part
\begin{equation}
\label{eq:H0}
	{\cal H}_0 (Q,t) = A(\Delta Q)^2+B(\Delta Q)^4- m p^*\Delta Q, 
\end{equation}
which does not depend on the external field, and then a second part creating the time-dependent linear chemical perturbation
\begin{equation}
\label{eq:Hpert}
	{\cal H}_\mathrm{pert}(Q,t) = m p(t)\Delta Q.
\end{equation}
By calculating the average values
\begin{equation}
\label{eq:U0a}
	U_0(t) = \int {\cal H}_0 (Q,t)\,P(Q,t)\, \mathrm{d}Q
\end{equation}
and
\begin{equation}
\label{eq:U0b}
	U_\mathrm{pert}(t) = \int {\cal H}_\mathrm{pert} (Q,t)\,P(Q,t)\, \mathrm{d}Q,
\end{equation}
respectively, we can thus divide the energy in its intrinsic and external contributions. For many experimentally relevant systems we can interpret them as mechanical ($U_0$) and chemical ($U_{\rm pert}$) contributions, respectively. For $Q$ being, for example, the particle size, the first one describes the elastic energy which changes over time only by variations of the particle distributions, while the second one depends directly on the time caused by the variable chemical product concentration $p(t)$.

\subsubsection{Duration of transient states}

We now estimate the duration of transient states for our 2-state model in the QS limit. We can call the transient state `activated' if its probability of occurrence is larger than the other, initial state.  We recognize that in the QS limit a minimum threshold of fueling concentration is needed to activate the transient state, given by the condition $p_{\rm max} \geq p^*$. This leads to the threshold (or `critical')  concentration for successful fueling 
\begin{equation}
\label{eq:ce0crit}
\begin{split}
	f_\mathrm{0, crit} &= \frac{p^*( k_{+}-k_{-})}{ k_{+}} \left[\left(\frac{ k_{+}}{k_{-}}\right)^\frac{k_{-}}{k_{-}- k_{+} }  - \left(\frac{ k_{+}}{k_{-}}\right)^\frac{ k_{+}}{k_{-}- k_{+}} \right]^{-1}\\
	&= p^*\left(1-\kappa\right)\left(\kappa^{\frac{\kappa}{1-\kappa}} - \kappa^{\frac{1}{1-\kappa}}\right)^{-1}
\end{split}
\end{equation}
which in the typical limit of $k_- \ll  k_+$ reduces to the simple relation $f_\mathrm{0, crit} = p^*$. As discussed above, $p^*$ signifies the important initial thermodynamic distance of the system to the transition point and can in principle be {\it a priori} designed. Once fixed, it directly defines the threshold concentration for successful fueling. We see that naturally also $p_{\rm sat}>p^*$ should hold for successful activation. 

 If we are above the threshold concentration for fueling, we can analytically estimate the duration of the times of the transient states.  This we can do in two ways:  
 
 {\it (i) Symmetry definition of the transient time:} In the case of a sufficiently high saturation limit, $p_{\rm sat}>p^*$, and the condition $p(t) >p^*$ holds, the stability bounds of the transiently stable state are very well defined by the times $t_1$ and $t_2 > t_1$ where $p(t)= p^*$, i.e., where the two states are equally likely.  (Note again that in our simple two-step chemical reaction the condition $p(t)= p^*$ is met only twice, during on-fueling and decay).  The duration of the transient state can then be formally defined as 
\begin{equation}
\label{eq:trans}
\tau_{\rm trans} = [t_2-t_1]_{p=p^*}
\end{equation}
where the notation means that the two times are evaluated if $p(t)=p^*$. We find through the {\it slow-decay} approximation $e^{-k_{-}t_1}\approx 1$, that
\begin{equation}
	t_1 = -\frac{1}{ k_{+}} \ln \left(1-\frac{p^*( k_{+} -k_{-} )}{ k_{+}f_\mathrm{0}} \right), 
\end{equation}
and the {\it fast-fueling} assumption $e^{-k_{+}t_2}\approx 0$, that
\begin{equation}
\label{eq:T_transient}
	t_2 = \frac{1}{k_{-}}\ln \left(\frac{ k_{+}f_\mathrm{0}}{p^*\left( k_{+}-k_{-} \right)} \right), 
\end{equation}
the duration of the transient state is essentially (and not surprisingly) determined by the on and off-rates of fueling, while there are logarithmic corrections depending on all rates and densities $f_0$ and $p^*$. 

Interestingly, we can show that for a bistable Hamiltonian of form (1) the times defining the transient states at $p\left(t \right) = p^*$ are extrema of  thermodynamic state functions, such as the free energy. Taking the derivative of the partition function $Z$ with respect to $p$, we find %\cmt{Sven, please check equation in terms of $\Delta Q$ and $A,B$. Please check consistently for the whole manuscript and  SI.}
\begin{equation}
\begin{split}
	\frac{\mathrm{d}Z}{\mathrm{d}p} &= -\int\limits_{-\infty}^\infty mQ\,e^{-\left(AQ^2 + BQ^4 + m\left(p\left(t \right) -p^* \right)Q\right)} \mathrm{d}Q.\\
\end{split}
\end{equation}
Obviously, the integrand becomes an antisymmetric function if $p\left(t \right) = p^*$ so that the integral vanishes and $Z$ has an extreme point, more precisely, leading to a maximum of the free energy $F(t)=-k_\mathrm{B}T\ln Z(t)$. Using the modified Bessel functions of the second kind $K_\alpha\left(x \right)$, we can write the extremum of the free energy:
%\begin{equation}
%\label{eq:Z_before}
%\begin{split}
	%Z_\mathrm{sym} &= \int\limits_{-\infty}^\infty e^{-AQ^4 - BQ^2} \mathrm{d}Q\\
	%%&= 2\int\limits_{0}^\infty e^{-AQ^4 - BQ^2} \mathrm{d}Q\\
	%%  &= e^{\frac{B^2}{4A}}\frac{1}{A^{1/4}} \int\limits_{\frac{B}{2\sqrt{A}}}^\infty e^{-u^2} \frac{1}{\sqrt{u - \frac{B}{2\sqrt{A}}}} \mathrm{d}u
	%&= e^{\frac{B^2}{8A}}  \frac{K_\frac{1}{4}\left(\frac{B^2}{8A} \right)}{2\sqrt{-\frac{A}{B}}}
%\end{split}
%\end{equation}
\begin{equation}
\label{eq:F_sym}
\begin{split}
	F_\mathrm{max} = F(p^*)	&= -k_\mathrm{B}T \left[\frac{A^2}{8B} +\ln\left( \frac{K_\frac{1}{4}\left(\frac{A^2}{8B} \right)}{2\sqrt{-\frac{B}{A}}}\right) \right]
\end{split}
\end{equation}
which in the QS limit constitutes the maximum work the system can perform.% \cmt{I think we should write $F_{\rm max}$. Please change if ok. Check if $A,B$ are correct according to the definition in the Hamiltonian. } 

{\it(ii)  Plateau definition of the transient time:} In the case of strong saturation (that is, small $p_{\rm sat}<p_{\rm max}^{\rm (us)}$), we obtain a longer period of a plateau behavior, for which $p(t)\simeq p_{\rm sat}=p_{\rm max}$.  Then the duration period of the transient state is mostly given by the time spent in the plateau.  This we can estimate by the following: at the plateau we have $p\simeq p_{\rm sat}=$~constant and thus $\dot{p} = \dot f -k_-p \simeq 0$, cf. eq.~(2),  and a linear decrease of the remaining fuel $f(t) \simeq f_0  - k_- p_{\rm sat}t $. The plateau decays when most of the fuel is consumed $f(t)\ll f_0$. Hence, we find for the duration of the plateau approximately  
\begin{equation}
\label{eq:plat}
	\tau_{\rm trans} \propto \tau_{\rm plateau} \simeq \frac{f_0}{k_{-}p_\mathrm{sat}}, 
\end{equation} 
constituting a useful law in dependence of the initial fuel concentration, decay rate, and saturation concentration. Noteworthy the transient time is now simply linear in $f_0$, if a plateau (i.e., saturation) behavior dominates the system. Due to the approximations made, however, a constant offset in this formula is generally plausible.

Note also that for saturating systems, the symmetry definition (i) should lead to values very close to the plateau definition (ii). The symmetry definition is more general and holds also for non-saturating systems, while for both $p>p^*$. Only the plateau regime (ii) leads to the clear linear scaling given by eq.~(\ref{eq:plat}).  

\begin{table*}
	\centering
		\begin{tabular}{l|c|c}
			{} & unit & description \\
			\hline
			$A$ & $\,k_\mathrm{B}T\mathrm{nm}^{-2}$& bimodal energy landscape parameter\\
			$B$ & $\,k_\mathrm{B}T\mathrm{nm}^{-4}$& bimodal energy landscape parameter\\
			$Q_\mathrm{c}$ & $\mathrm{nm}$ & center of unperturbed energy landscape\\
			$m$ & $k_\mathrm{B}T\,\mathrm{l}\,\mathrm{mmol}^{-1}\mathrm{nm}^{-1}$& impact of $p$ on the energy landscape\\
			$\tilde k_{+}$ & $\mathrm{l}\,\mathrm{mmol}^{-1}\mathrm{h}^{-1}$ & $p$-forming rate\\
			$k_{-}$ & $\,\mathrm{h}^{-1}$ & $p$-decomposition rate\\
			$p_\mathrm{sat}$ & $\mathrm{mmol}\,\mathrm{l}^{-1}$ & saturation value for $p$\\
			$p^*$ & $\mathrm{mmol}\,\mathrm{l}^{-1}$ & initial skew / transient threshold\\
			\hline
			$f_0$ & $\mathrm{mmol}\,\mathrm{l}^{-1}$ & initial fuel concentration (taken from experiment)
		\end{tabular}
	\caption{Model parameters with corresponding units for the fits of the experimental data of Heckel {\it et al.}~\cite{Heckel_2021} presented in Fig.~\ref{fig:Fits_global}. $f_0$ is input from experiments, while the others are global or $pH$-dependent fitting parameters, see text and Table II. } %\cmt{ATTENTION: units are not consistent with the next table.}}
	\label{tab:fitparameters}
\end{table*}

\subsection{Slow system relaxation: Smoluchowski approach}
If the system relaxation is slow compared to the chemical reaction, its response to the time-dependent Hamiltonian will be retarded. As a simple start, we can assume the system is following an overdamped diffusive dynamics following the Smoluchowski (drift-diffusion) equation~\cite{doi,risken}: 
\begin{equation}
	\frac{\partial P}{\partial t} = D \frac{\partial^2 P}{\partial Q^2} - D \beta \left[\frac{\partial^2 {\cal H}}{\partial Q^2} P +\frac{\partial {\cal H}}{\partial Q} \frac{\partial P}{\partial Q} \right]
	\label{eq:smol}
\end{equation}
where $P=P(Q,t)$ is the time-dependent probability distribution, $D=k_\mathrm{B}T/\xi$ the diffusivity which we assume $Q$-independent, and $\xi$ the friction coefficient. We solve this equation numerically using the {\it fplanck} python package~\cite{Holubec_2019}. Note that time-dependent averages in the system can be still evaluated with general eq.~(\ref{eq:P}). 

For an interpretation of the results during slow dynamics we need to briefly discuss the timescales in this problem: 
A diffusion timescale can now be defined by {$\tau_\mathrm{D} = \left(Q_2 - Q_\mathrm{c}\right) / D$, where $Q_2$ is the position of the initial global minimum of ${\cal H}{(Q)}$. Note, that this position changes slightly over time as the linear term of the Hamiltonian does. For our definition, we thus use the symmetric situation at $p\left(t \right) = p^*$ where it holds that $\tau_\mathrm{D} = -\frac{A}{2BD}$. This time expresses how long}
%, that is, the time
the system needs to diffuse from the minimum to the barrier position.

Such a diffusive time can be readily related to the typical fueling time through the dimensionless parameter 
\begin{equation}
\label{eq:alpha}
\alpha = \tau_\mathrm{D} k_+ 
\end{equation}
If $\alpha \ll 1$, we are in the QS limit. For $\alpha \gtrsim 1$ the system relaxes slow and significantly retarded to the chemical reaction. In the extreme case of $\alpha \gg 1$, the system never relaxes during fueling and essentially does not change in time. 

Another relevant timescale in this bistable system, if energy barriers are significant ($\Delta {\cal H} \gtrsim k_\mathrm{B}T$), is the so-called Kramers time for diffusive barrier crossing~\cite{Kramers_1940,RevModPhys.62.251}. There are several expressions for it depending on the approximations made. We define the Kramers time with $\tau_D$ as a prefactor to the important exponential Arrhenius-factor to have it consistent in the vanishing barrier limit, hence, 
%\begin{equation}
	%\tau_{\rm Kramers} = 2\pi\xi \left[{\cal H}''\left(Q_2 \right) \left|{\cal H}''\left(Q_c \right) \right| \right]^{-1/2} e^{\beta {\Delta H}}
	%\label{eq:Kramers}
%\end{equation}
\begin{equation}
	\tau_{\rm Kramers} = \tau_\mathrm{D} e^{\beta {\Delta {\cal H}}},
	\label{eq:Kramers}
\end{equation}
with energy barrier $\Delta {\cal H}$. %
%, and we write $Q_2,Q_c$ to denote formally the positions of the stable minimum and the transition state, respectively.
For large barriers, the Kramers time is limiting for the distribution to flood the metastable state once activated by the fuel, and also for the reverse process. Note that $\Delta {\cal H}(t)$  as well as the location of the extrema are themselves time-dependent, hence the barrier crossing effects are not uniquely to quantify. For simplicity, we follow the rule that we calculate the Kramers time using $\Delta {\cal H} = {\cal H}(Q_c)-{\cal H}(Q_2)$ from the symmetric, non-skewed energy landscape at $p=p^*$. The ratio between Kramers time and the chemical fueling time, we denote then by $\alpha_\mathrm{K} = \tau_{\rm Kramers}k_+$.

\section{Case study: chemical fueling of hydrogel collapse}

We now apply our model in the QS limit to explicitly fit the experimental data of chemical fueling of hydrogel collapse recently put forward by Heckel {\it et al.}~\cite{Heckel_2021}. In these experiments, the chemical fuel N-Ethyl-N'-(3-dimethy- laminopropyl)carbodiimide (EDC) was used to trigger the volume phase transition (VPT) for poly(methacrylic acid) (PMAA) microgels, to demonstrate that the collapsed hydrophobic state can be programmed in time using the fuel concentration in a cyclic reaction network. The EDC addition enables two neighboring carboxylic acid groups to form a cyclic carboxylic anhydride which increases the hydrophobicity of the hydrogel.  The measured observable was the finite hydrogel radius $R(t)$, see Fig.~3, averaged over many particles for fixed time $t$. We assume $R(t) = \langle Q(t) \rangle$ as a an ensemble average over a hypothetically infinitely large sample of spheres. 

\begin{figure*}
	\centering
		\includegraphics[width=0.7\textwidth]{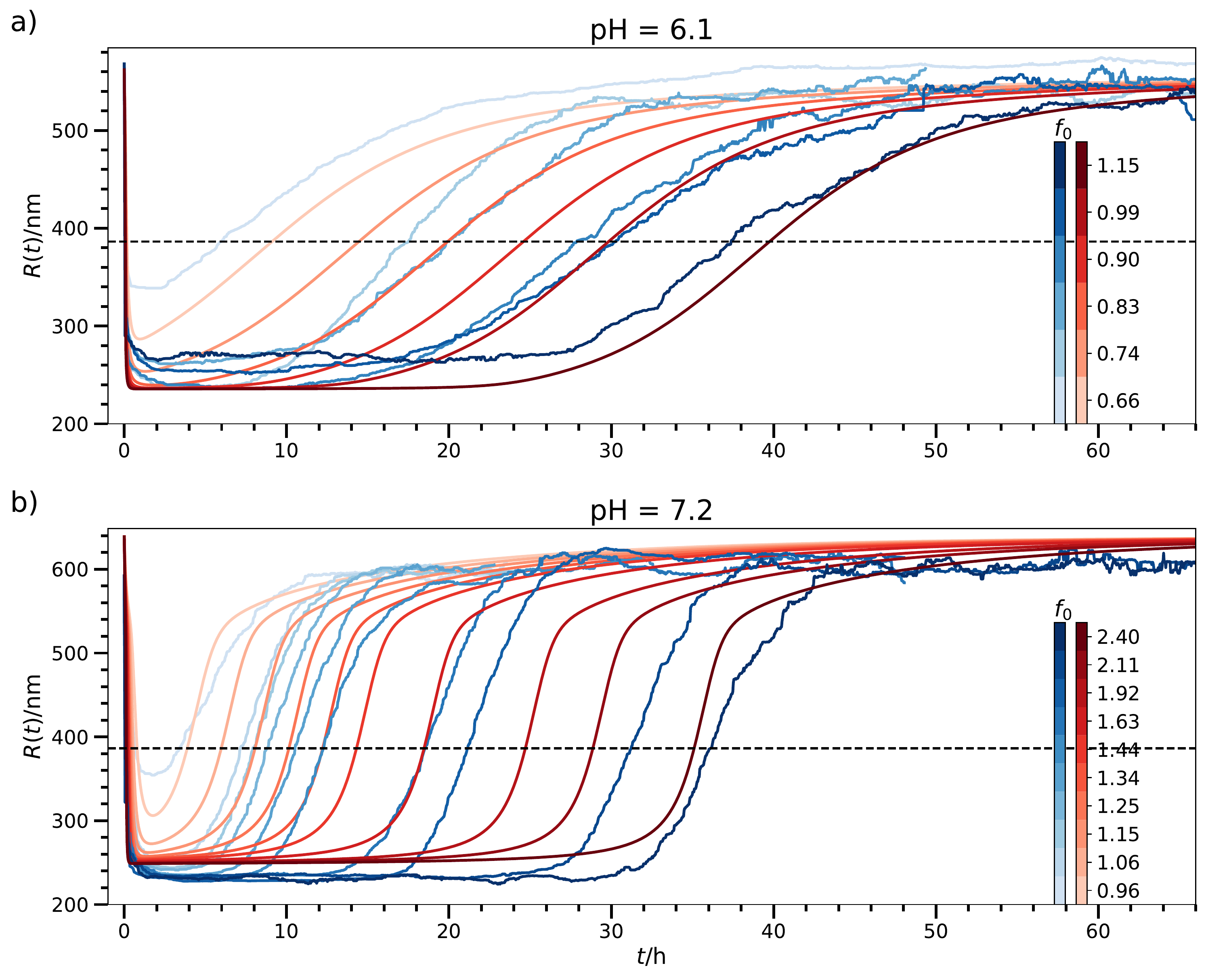}
	\caption{Model fits for a fixed energy barrier of $2\,k_\mathrm{B}T$ to the data set of chemically fueled hydrogel collapse~\cite{Heckel_2021} in the QS limit for (a) pH=6.1 and (b) pH=7.1. The color bar codes the different fueling concentrations $f_0$. % in units of mmol/l which we assume to be the same as the equivalent ECD concentration in the experimental work.
	The blueish hues represent the experimental and the reddish ones the theoretical data. {The horizontal dashed line indicates the mean radius at the symmetric state, $R(p^*)$, below which the transient state is activated. {Note that in contrast to the original work by Heckel {\it et al.}~\cite{Heckel_2021} we choose $f_0$ in units of mmol/l, but not in `equivalent EDC' which is stoichmetrically scaled by twice the concentration of carboxylic acid groups. These units can be converted via $1\,\mathrm{equiv.\,\,EDC} = 0.96\,\mathrm{mmol/l}$}. }}
	\label{fig:Fits_global}
\end{figure*}

\subsection{Fitting procedure}

Our fit is based on the numerical solution of the coupled eqs.~(\ref{eq:Hm}) and (\ref{eq:c5}) including saturation.  The EDC fuel concentration is translated to $f_0$ in units of $1\,\mathrm{mmol/l}$.  {In Heckel {\it et al.} we find several measurement curves for two pH-values,  pH = 6.1 and pH = 7.2, with differing initial fuel concentration $f_0$. We reconcile our model and the experimental data by minimizing the total mean squared deviations (MSD) between theory curves and experimental data for a fixed pH (details in the Supporting Information, ESI). The model takes the initial fuel concentration $f_0$ as an input, while the other parameters listed in Table~\ref{tab:fitparameters} are fitting parameters, now with real experimental units assigned. {\it However, all of the fitting parameters are kept constant for a fixed pH value.} When a fitting parameter is called {\it global}, it means that it is constant also for both pH values. 

We treat the parameters for the energy landscape $A$, $B$ and $Q_\mathrm{c}$ as global parameters, because we expect them to be intrinsic properties of the hydrogel colloids, which do not depend on the pH-value. We also globally fix $p_\mathrm{sat}$, which should be constant because pH-dependent side reactions changing the amount of microgels are not included in our model. In contrast, the chemical reaction rates \cite{Kariyawasam_2020, Williams_1981} and also the $m$-value describing their impact on the energy landscape may depend on the pH-value. (We note that removing the constraints of pH-independent energy landscape parameters we obtain slightly improved fits, but because the improvement is only little (see ESI Fig.~S\ref{fig:Fit2kBT_pH}), we consider a pH-dependent $m$-value as sufficient. 

Moreover, we observe some arbitrariness (i.e., some insensitivity) of the fits regarding the precise magnitude of the energy barrier.  We avoid this problem by fixing the energy barrier for a fixed pH through the exact expression in the symmetric landscape, $\Delta {\cal H} = \frac{A^2}{4B}$.  Interestingly, unimodal potentials of the simple form $\propto \Delta Q ^{2n}$ with $n=1,2$ were not able to reproduce the relatively fast sigmoidal transitions from one state to the other ({see Supporting Information Fig.~S\ref{fig:Quadratic} and S\ref{fig:Q6}}).  {Improved fits were achieved by broader $n=3$ and a square-well `box' potential. But here the distribution functions become relatively broad, in contrast to the experiments~~\cite{Heckel_2021}, with unrealistically unbound values of the radius $R$. Hence, a Landau-like quartic potential including the presence of transition barriers ($\Delta {\cal H} \gtrsim k_\mathrm{B}T$) was most adequate to fit the data. Note that hydrogel charge content tunes the location and width of the VPT~\cite{Emanuela}.}

Further modifications of the fitting constraints are conceivable. For example, we can pre-fix $p_\mathrm{sat}$ to the number of available reaction partners, allow an offset for $f_0$, or change pH-dependent variables to global ones.  Similarly, we tested purely unsaturated equations for $p$, eq.~(3),  but dropped them due to the following reasons. Without saturation, the maximum value of $p$ is not bounded but can exceed the (experimentally roughly known) number of reaction partners. In addition, the pronounced peaks in $p(t)$ without saturation make it hard to reproduce the flat plateaus we observe in $R\left(t \right)$ for large $f_0$ (see Fig. \ref{fig:c_comp}). Finally, without saturation  the fast conversion to $p$ and thus enhanced fuel consumption leads to a sublinear scaling between $f_0$ and the transient time, which is in contrast to the fully linear experimental observation, cf. Fig.~\ref{fig:tau} later. 

%We use curve fit from scipy.optimize with the leastsq method as fitting routine~\cite{xx}. We use this plot as a reference, since it generates a realistic energy landscape with appropriate barrier height...
\begin{table}[h!]
	\centering
		\begin{tabular}{l|c|c}
			{} & $\mathrm{pH} = 6.1$ & $\mathrm{pH} = 7.2$\\
			\hline
			$A\,\left[k_\mathrm{B}T\,\mathrm{nm}^{-2}\right]$	&	\multicolumn{2}{c}{$-1.89 \times 10^{-4}$}\\
			$B\,\left[k_\mathrm{B}T\,\mathrm{nm}^{-4}\right]$	&	\multicolumn{2}{c}{$4.45 \times 10^{-9}$}\\
			$Q_\mathrm{c}\,\left[\mathrm{nm}\right]$	&	\multicolumn{2}{c}{$386$}\\
			$m\,\left[k_\mathrm{B}T\,\mathrm{l}\,\mathrm{mmol}^{-1}\mathrm{nm}^{-1}\right]$	&	$6.87 \times 10^{-2}$	&	$0.260$\\
			%$f^*\,\left[\mathrm{mmol}\,\mathrm{l}^{-1}\right]$	&	\multicolumn{2}{c}{$0.00$}\\
			$\tilde k_{+}\,\left[\mathrm{l}\,\mathrm{mmol}^{-1}\mathrm{h}^{-1}\right]$	&	$25.5$	&	$11.9$\\
			$k_{-}\,\left[\mathrm{h}^{-1}\right]$	&	$2.06 \times 10^{-2}$	&	$5.79 \times 10^{-2}$\\
			$p_\mathrm{sat}\,\left[\mathrm{mmol}\,\mathrm{l}^{-1}\right]$	&	\multicolumn{2}{c}{$0.799$}\\
			$p^*\,\left[\mathrm{mmol}\,\mathrm{l}^{-1}\right]$	&	$0.550$	&	$0.749$\\
			\hline
			$k_{+} = \tilde k_{+}p_\mathrm{sat}\,\left[\mathrm{h}^{-1}\right]$	&	$20.4$	&	$9.53$\\
			$Q_1 = Q_\mathrm{c} - \sqrt{-A/2B}\,\left[\mathrm{nm}\right]$	&	\multicolumn{2}{c}{$241$}\\
			$Q_2 = Q_\mathrm{c} + \sqrt{-A/2B}\,\left[\mathrm{nm}\right]$	&	\multicolumn{2}{c}{$532$}\\
			
		\end{tabular}

	\caption{Values of the resulting parameters for an optimal fit of the data for transient hydrogel fueling plotted in Fig.~\ref{fig:Fits_global}. $A, B, Q_\mathrm{c}$, and $p_{\rm sat}$ are global, pH-independent fitting parameters. {The last line separates $k_{+}$, $Q_1$ and $Q_2$ which are characteristic for our energy landscape but directly depend on some of the fitting parameters.}}  % \cmt{ATTENTION: units are not consistent with the previous table. I would keep the nm for size, h for time, and mmol/l for concentrations. For A and B you can choose. If number are of the order of $10^-1$ - $10^2$,  I would write them out with maximum 3 digits. Otherwise use the $10^n$ form.}}
	\label{tab:fit_results}
\end{table}

\subsection{Fitting results}

The results of an exemplary best fit are displayed in Fig.~\ref{fig:Fits_global}. Here, we fixed the energy barrier to 2~$k_\mathrm{B}T$. We obtain comparable results using other energy barriers (see Fig.~S\ref{fig:Fit5kBT} for $\Delta {\cal H} = 5\,k_\mathrm{B}T$), which confirms that the exact choice of the barrier height is of minor importance in the QS case, as long as is it not vanishing ($\Delta {\cal H} \gtrsim 1 \,k_\mathrm{B}T$). Later we will see, however, that the precise value of $\Delta {\cal H}$ makes a substantial difference in the full nonequilibrium when the system relaxation is comparable to the chemical reaction times.  

The parameters of this fit are summarized in Table~\ref{tab:fit_results}. The fits themselves are not quantitative, but considering the relatively large error in the time domain of the experiments of a few hours (cf. Fig. 2b in ~\cite{Heckel_2021}), they are satisfactory, in particular, they agree very well in several qualitative aspects: They describe the fast drop of $R(t)$ in the experiments {down to $236\,\mathrm{nm}$ and $224\,\mathrm{nm}$ for pH~$ = 6.1$ and 7.2, respectively (compared to the values obtained by our fits: $236\,\mathrm{nm}$ and $249\,\mathrm{nm}$),} the plateau behavior, the slow rise of the radius, and, in particular, the trends and magnitudes of the transient times, as in detail discussed later. 
%Max values:
%Experiments: $575\,\mathrm{nm}$ and $625\,\mathrm{nm}$
%Fits: $562\,\mathrm{nm}$ and $639\,\mathrm{nm}$

Note that our model fits allow to reversely calculate the time evolution of fuel $f(t)$ and/or products $p(t)$ in the system. For the parameter set of Fig.~\ref{fig:Fits_global} and the exemplary choices pH~$=6.1$ and $f_0 = 0.99\,\mathrm{mmol/l}$, we show the temporal evolution of the products $p(t)$ together with the corresponding time evolution of the energy landscape ${\cal H}(Q,t)$ in Fig.~\ref{fig:c}. One can nicely compare the features of products and mechanical response at different characteristic times, including the saturation behavior and the symmetric states at $p=p^*$. 

{More conclusions can be drawn from the fitting numbers in Table~\ref{tab:fit_results}.  Since $f_0/p_\mathrm{sat}\leq2.5$, we expect high $k_{+}/k_{-}$ ratios because of the pronounced plateaus. Indeed, our $k_{+}$ are about three orders of magnitude larger than $k_{-}$. This means, with respect to Sharko {\it et al.}~\cite{Sharko_2022}, we are moving through different fueling regimes depending on the initial fuel concentration $f_0$. In particular, our model captures the transition from small dips (minima) to ever-widening plateaus.} {Furthermore, we find that the expected pH-dependent reactivity changes with our parameters: When pH increases, we expect slower anhydride formation \citep{Kariyawasam_2020, Williams_1981} and faster hydrolysis \citep{Woodruff_1972}, which is expressed in smaller $k_{+}$ and lower $k_{-}$ for pH~$= 7.2$ than for $6.1$. This change in reactivity explains in turn, why we need more fuel for the same drop in radius for pH~$= 7.2$.} Moreover, we find excellent agreement between the saturation concentration $p_\mathrm{sat}$ when compared to the experimental numbers~\cite{Heckel_2021}.} 
%
%{Moreover, we find excellent agreement between the saturation concentration $p_\mathrm{sat}$ and its experimental counterpart. Actually, it should be $1\,\mathrm{eq.\,EDC} = 0.96\,\mathrm{mmol}\,\mathrm{l}^{-1}$, but as described in Heckel {\it et al.}~\cite{Heckel_2021, Berg-Feld_1982}, we have an upper anhydrid formation limit of $83\,\%$ due to statistical reasons, which matches perfectly with our value. \cmt{I do not understand. Wich value? } We also tried fits, where this parameter \cmt{Which parameter?} is not free but fixed to the experimental value. Hereby, only the chemical reaction rates were changed slightly, but the general fits quality was almost not affected. \cmt{I do not understand well this paragraph, is it necessary? How is the 2nd sentence connected to the 1st one on saturation?}

Of special interest is the behavior of the duration of the transient times and how they are controlled by the physical and chemical parameters in the system. In the experimental paper ~\cite{Heckel_2021} the transient times were defined as {the timespans between $R^* = (R\left(t=0 \right) + R_\mathrm{min}) /2$, where $R_\mathrm{min}$ is the minimum radius for each individual curve. In the following, we call this time \textit{half collapse time}, denoted by $\tau_{1/2}$. This definition we can in principle also apply to the data generated by our model.}% In addition, we have for the model data the transient time defined by equation~\ref{eq:trans}.}
\begin{figure}[h]
	\centering
		\includegraphics[width=0.45\textwidth]{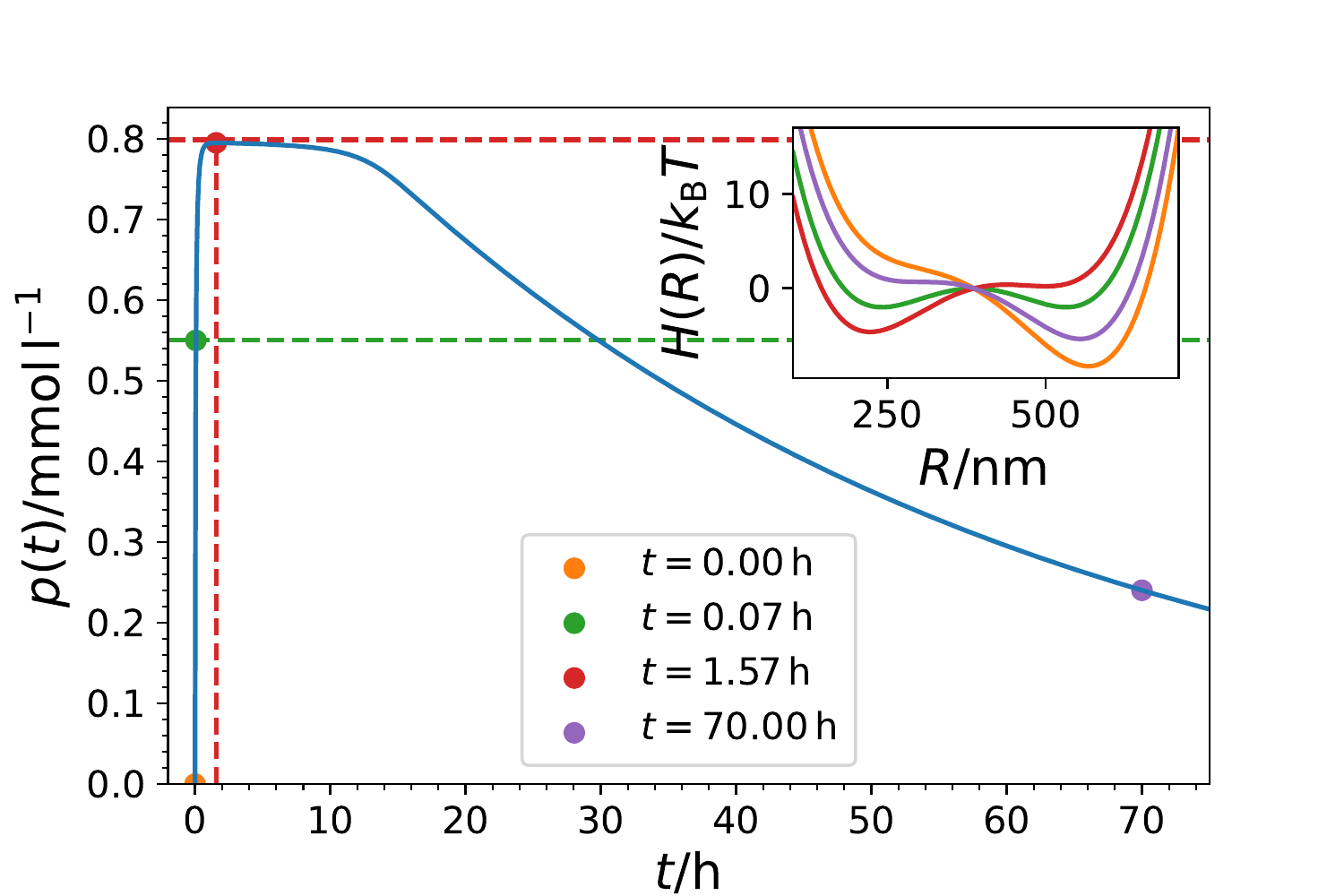}
	\caption{Model prediction of the temporal evolution of the product concentration $p(t)$ in the system of Heckel {\it et al.}~\cite{Heckel_2021}. {Parameters were taken from Table~\ref{tab:fit_results} (pH$ = 6.1$, $f_0 = 0.99$~mmol/l)}. The colored symbols depict different times for which the bimodal landscape ${\cal H}(Q)$ is plotted in the inset using the same color code. The values of $p_{\rm sat}$ and $p^*$ are indicated by horizontal red and green dashed lines, respectively. The time of the maximum $t_{\rm max}$, when $p(t_{\rm max})=p_{\rm sat}=p_{\rm max}$,  is indicated by the red vertical line.}%\cmt{We should mention the pH and parameters here, or refer to a Table.}}
	\label{fig:c}
\end{figure}

However, we evaluate the transient times, $\tau_{\rm trans}$, in our model according to our well-defined symmetry definition eq.~(\ref{eq:trans}), which is similar but not exactly the same as $\tau_{1/2}$. The definition is applied to the fitting curves in Fig.~3 and plotted vs. $f_0$ in Fig.~\ref{fig:tau}, in which we compare also to the experimental definition $\tau_{1/2}$. There is overall very good agreement. In particular,  this plot suggests a linear connection between $\tau$ and $f_0$, as predicted from our analysis of the transient times in the saturated plateau regime, eq.~(\ref{eq:plat}).
{Using the fitted values in Table~\ref{tab:fit_results} we obtain slopes of $1.01\cdot10^{5}$ and $3.59\cdot10^{4}\,\mathrm{h}\,\mathrm{l}\,\mathrm{mmol}^{-1}$ for pH=6.1 and 7.2, respectively. They are shown as straight lines in Fig.~\ref{fig:tau} where {the $y$-intercept is chosen consistently from our fits where $\tau\left(p^* \right) = 0$, i.e. where activation of the transient state starts.} %the offset is adjusted in order to minimize the squared errors compared to the experimental $\tau_{1/2}$.
Linear fits of  the experimental $\tau_{1/2}$ provide $9.83\cdot10^{4}$ and $3.70\cdot10^{4}\,\mathrm{h}\,\mathrm{l}\,\mathrm{mmol}^{-1}$ (dotted lines) underlining the agreement between experiment and theory. Hence, we understand the linear relation between $\tau_{1/2}$ and $f_0$ described by Heckel {\it et al.}~\cite{Heckel_2021} in a mathematical framework, which facilitates lifetime tuning of the transient state.} 

%The slope of the lines is $9.83\cdot10^{4}$ and $3.70\cdot10^{4}$ for pH=6.1 and 7.2, respectively. This compares consistently with the slopes calculated from $1/(k_- p_{\rm max})$ as from eq.~(\ref{eq:plat}) using the fitted values in Table \ref{tab:fit_results} which are $1.01\cdot10^{5}$  $3.69\cdot10^{4}$...

%\sven{Offset discussion: No response below critical value in experiment, continuous transition in theory, introducing an offset improves fit quality and diminishes the response below the critical value enormously, in combination with inertia we can explain the non response, buffering is maybe a chemical question, according to our theory, this offset does not change our analysis of $\tau_{1/2}$ since it affects only the offset but not the slope}
%\cmt{please complete discussion and conclusion}

%Hence, our model provides a deep and quantitative physical understanding of the experimental time scale of transient fueling. 

\begin{figure}[h]
	\centering
		\includegraphics[width=0.45\textwidth]{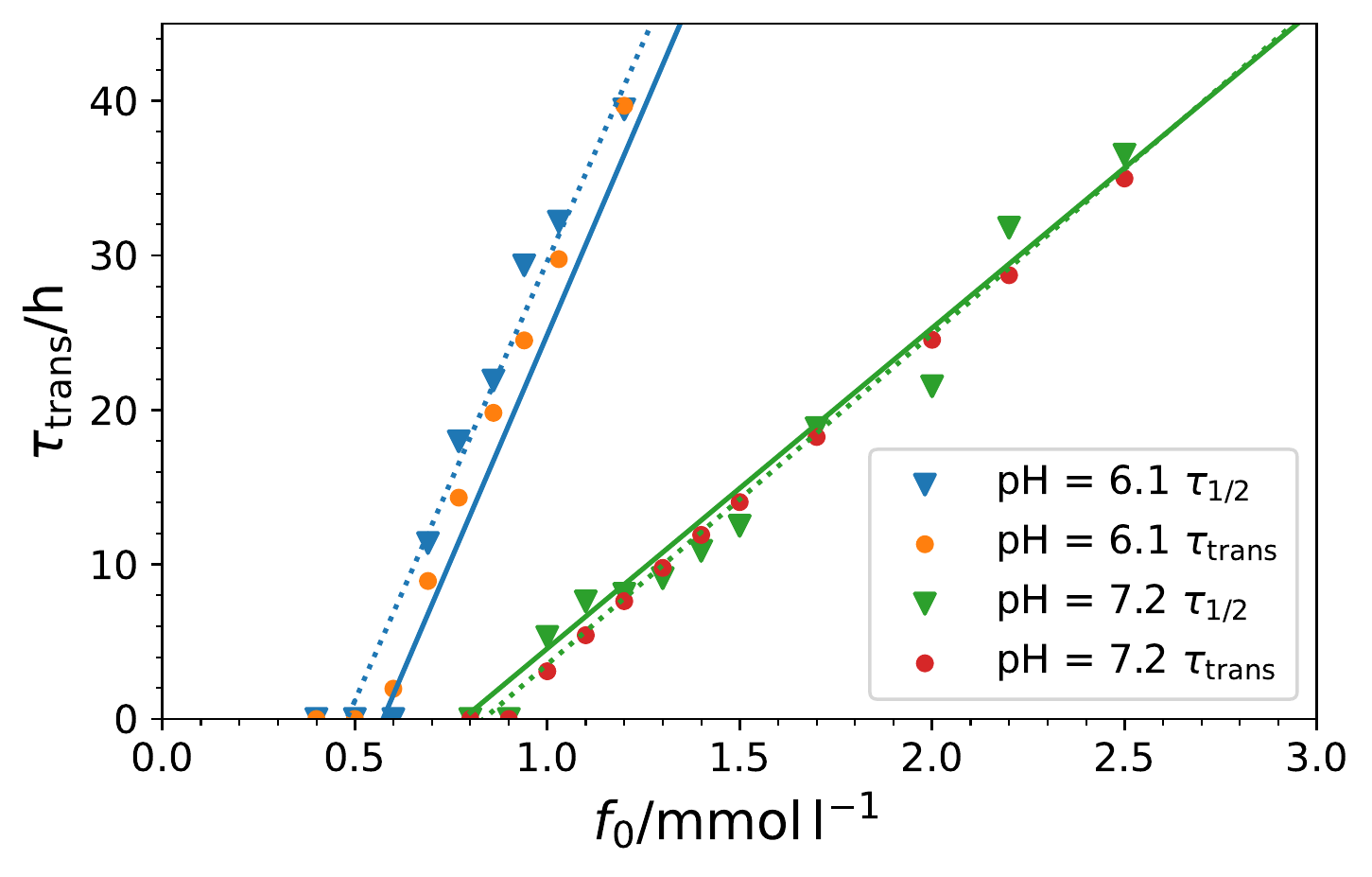}
	\caption{Duration times of transient states, $\tau_\mathrm{trans}$, eq.~({\ref{eq:trans}}), for both pH-values in dependency of the initial fuel concentration $f_0$ (solid circles) compared to the experimentally evaluated times $\tau_{1/2}$~\cite{Heckel_2021} (filled triangles, fitted linearly by the dotted lines). The solid lines show slopes  calculated by the theoretical `plateau regime' prediction in the  eq.~(\ref{eq:plat}), with $y$-intercept fixed consistently with $\tau(p^*)=0$ with $p^*$ from Table~II.}%The lines were shifted in vertical direction in such way that the least squares related to the experimental $\tau_{1/2}$ are minimized, i.e., agree best within an offset. \cmt{Do we need this?}}
	\label{fig:tau}
\end{figure}

\subsection{Thermodynamic analysis in the QS limit}

\begin{figure}[h]
	\centering
		\includegraphics[width=0.45\textwidth]{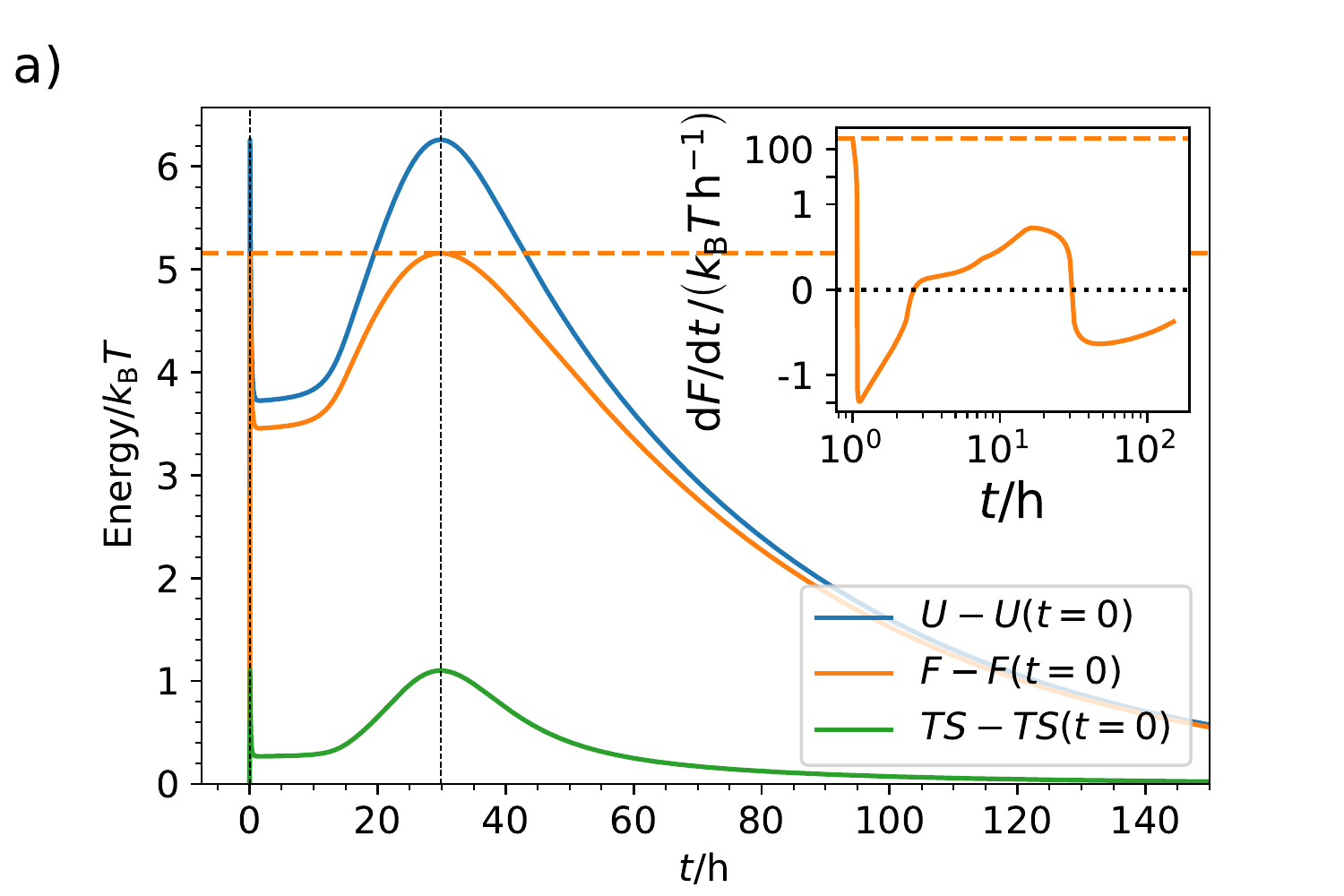}\\
		\includegraphics[width=0.45\textwidth]{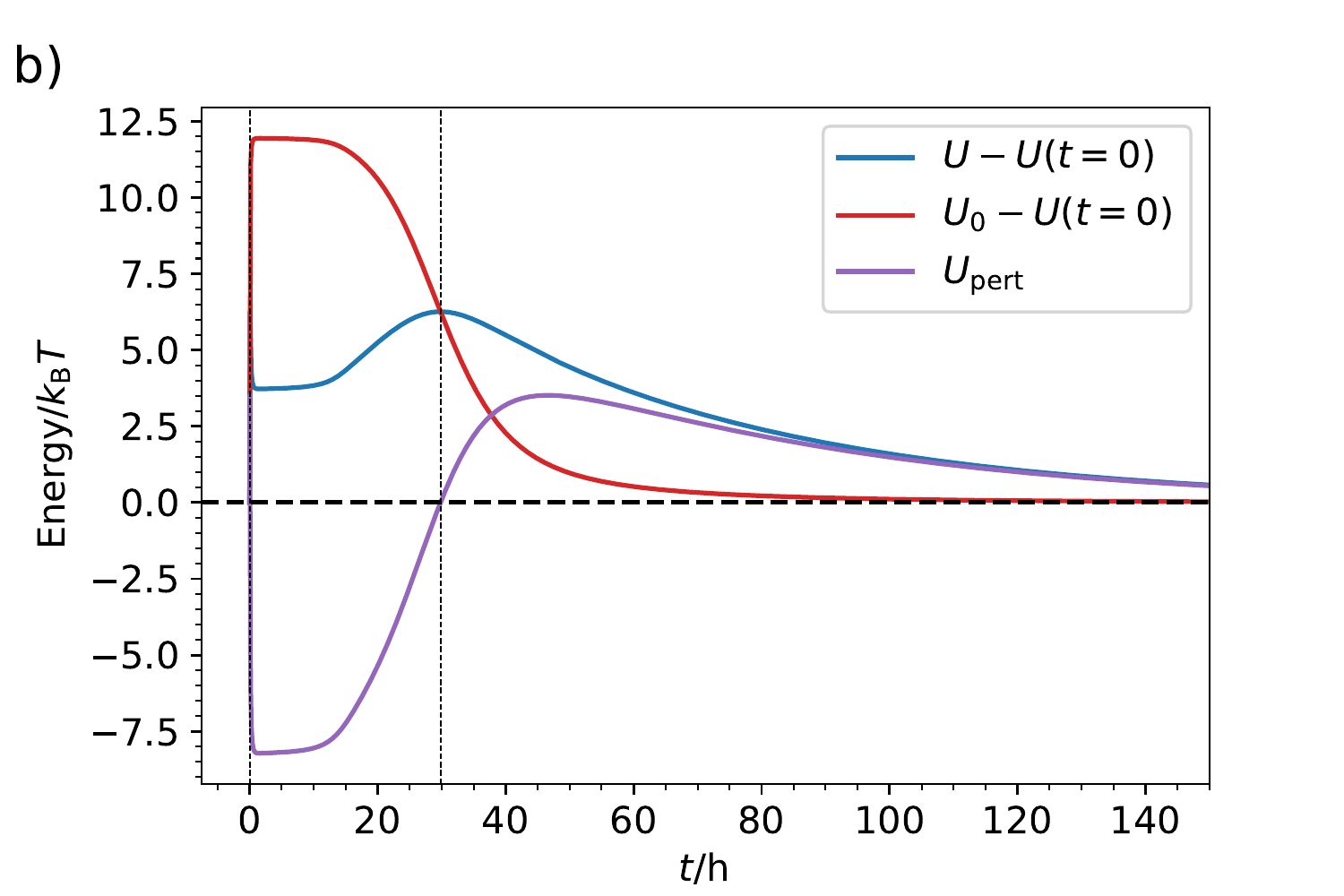}
	\caption{(a) Temporal evolution of the energy $U(t)$, free energy $F(t)$ and entropic contribution $TS(t)$ subtracted by their initial values at time $t=0$ for $f_0=0.99$~mmol/l and parameters taken from Table~\ref{tab:fit_results} at pH~=~6.1. The maximum of $F$ given by eq.~(\ref{eq:F_sym}) is marked by the horizontal dashed (orange) {while the vertical dotted lines mark the two times where $p=p^*$}. The inset shows the power (work per time, $P=\mathrm{d}F/\mathrm{d}t$) as a function of $t$ {(with time axis shifted by $1\,\mathrm{h}$ to allow for a logarithmic scale. {Because of negative values of $P$, the scale on the $y$-axis is linear between $-0.01$ and $0.01\,k_\mathrm{B}T\mathrm{h}^{-1}$ and logarithmic elsewhere. For the negative part, we show $-\log\left(-P\right)$}). {The horizontal dashed line shows the analytical limit of eq.~(\ref{eq:power}) at $t=0$.} (b) The total energy $U$ split in mechanical ($U_0$) and chemical contributions ($U_{\rm pert}$) according to eqs.~(\ref{eq:U0a}) and (\ref{eq:U0b}).}  }
	\label{fig:Therm}
\end{figure}

In the QS limit, thermodynamic quantities such as energy, entropy, etc., are time-dependent but still well defined. For selected parameters (cf. Table~\ref{tab:fit_results}, at pH~=~6.1, $f_0=0.99$~mmol/l) we show $U(t)$, $S(t)$ and the free energy $F(t)$ in Fig.~\ref{fig:Therm}.  {We have already discussed $F(p)$ and found that it has a maximum at $p\left(t\right)=p^*$ where the Hamiltonian is symmetric and which is realized two times in the system: at very short times ($t\simeq 0$ on this scale) where $F(t)$ jumps up and down in a $\delta$-peak like fashion and again at about $t\simeq 30$~h. (Note that if we do not reach the transient state because $p_\mathrm{max}<p^*$, then $F$ would only have one maximum.) The behavior of $F(t)$ looks complex, since $F$ drops to a local minimum between the two maxima, at which we have the transient plateau behavior, during which $p \simeq p_{\rm sat} = p_\mathrm{max}$. After the second maximum $F(t)$ relaxes back to the initial state. The apparently complexity essentially arises from the mapping of the asymmetric chemical kinetics $p(t)$ on the behavior of $F(p)$ according to the bistable Hamiltonian, eq.~(1).  As one can see in Fig.~\ref{fig:Therm}(a), $U$ and $S$ have similar functional forms than $F$.  For an unimodal Hamiltonian less complex behavior can be expected. 

The inset of Fig.~\ref{fig:Therm}(a) shows the derivative of the free energy, $P=\mathrm{d}F/\mathrm{d}t$, which we can interpret as the thermodynamic power. It has its largest absolute values immediately at the addition of the fuel at $t=0$, meaning that the system is most active initially.  This makes sense given the high $k_{+}$ to $k_{-}$ ratio. It can also be understood from the analytical results, eq.~(\ref{eq:power}) which is largest at the beginning where the radius is biggest and products massively produced, $P(t=0) \simeq m f_0 k_+ (Q_2-Q_\mathrm{c})$. For later times during relaxation then broader and flatter peaks in the power develop at around $t\simeq 30$~h, when $F(t)$ peaks again and the system transitions back to the initial state. Hence, most work is performed chemically and elastically at the transitions to and from the transient state in a bistable system. %\cmt{Sven, please check}

Part b) of Fig.~\ref{fig:Therm} displays the course of the total energy $U(t)$ and its mechanical and chemical contributions, $U_0$ and $U_{\rm pert}$, respectively. The most intuitive is the mechanical energy $U_0$. By addition of the fuel we force the hydrogel very quickly to the collapsed state whereby energy is stored (increased) elastically. Its time of maximum coincides with that of maximum products, $p=p_\mathrm{sat}$, and then we observe relaxation, where the stored energy is released again. For the total $U$, the shape is different. Here, we observe the maxima as in $F(t)$ where $p=p^*$, otherwise the energy is always lower. To understand this, let us consider the chemical (or external) part $U_\mathrm{pert}$, being the difference of $U$ and $U_0$: It increases $\delta$-peak like in the very beginning ($t\simeq 0$ on this scale) when chemical energy is quickly pumped into the system and converted to mechanical one. When the mechanical energy starts relaxing, $U_\mathrm{pert}$ rapidly decreases and turns negative. Subsequently, the process is reversed but not chemically fueled, driven by the stored elastic energy. Note that $U_\mathrm{pert}$ has zeros at short times $p\left(t\right) \simeq 0$ (on this scale) and at $p\left(t\right) = p^*$. This is accompanied with maxima in the entropy, cf. Fig.~6(a), indicating heat exchange with the bath along the evolution of the internal energy.}

%\begin{figure}[h]
	%\centering
		%\includegraphics[width=0.45\textwidth]{Plots/Therm.pdf}
	%\caption{Thermodynamic quantities such as internal energy $U$, entropy $S$ and free energy $F$ in the QS limit. \cmt{We should mention the pH and parameters here, or refer to a Table.}}
	%\label{fig:Therm}
%\end{figure}
%$k_{+}$ to k_{-} ratio
%\begin{figure}[h]
	%\centering
		%\includegraphics[width=0.45\textwidth]{Plots/U_sep.pdf}
	%\caption{\sven{Energy separated in mechanical and chemical contributions}}
	%\label{fig:Therm_sub}
%\end{figure}

%\begin{figure}[h]
	%\centering
	%\begin{subfigure}[b]{0.45\textwidth}
         %\centering
		%\includegraphics[width=1\textwidth]{Plots/U_sub_a.pdf}
		%\caption{}
         %\label{fig:Therma}
     %\end{subfigure}
		%
		%\begin{subfigure}[b]{0.45\textwidth}
         %\centering
		%\includegraphics[width=1\textwidth]{Plots/U_sub_b.pdf}
		%\caption{}
		%\label{fig:Thermb}
     %\end{subfigure}
		%
	%\caption{\sven{Left: Temporal evolution of thermodynamic quantities, Right: $U$ splitted in mechanical and chemical contributions}}
	%\label{fig:Therm}
%\end{figure}

\subsection{Effects of slow system relaxation}

We finally discuss the effects of slow relaxation, {such as possible delays in the system's response \cite{Fusi_2023}, in the context of the just discussed case of chemically fueled hydrogel collapse. For this, we fix the parameters according to {Table~\ref{tab:fit_results}, for pH=6.1 and $f_0=0.99\,\mathrm{mmol/l}$}. We now tune the timescale separation parameter $\alpha$ in eq.~(\ref{eq:alpha}) from {$0.1$ to $10$} and solve numerically the Smoluchowski equation (\ref{eq:smol}) for the distribution $P(Q,t)$ to calculate averages, such as $R(t)$. The results are shown in Fig.~\ref {fig:RSmol}(a).  For $\alpha=0.1$ we are still very close to the QS limit because the diffusive system relaxation is still 10-fold faster than the fast rate $k_+$. However, moving to $\alpha=0.5$ or 1, we observe clear retardation effects, involving less collapse and a delay of the minimum in time. For  $\alpha=5$ and larger, the system becomes quite inert and the response effects very small. The chemical powers transform mostly into dissipative losses and cannot be used to perform work. Clearly, a change of internal timescales can change the nonequilibrium time evolution and thermodynamics massively.   

\begin{figure}[h]
	%\centering
	%\begin{subfigure}[b]{0.45\textwidth}
         %\centering
		%\includegraphics[width=1\textwidth]{Plots/R_beide_DUklein.pdf}
		%\caption{}
         %\label{fig:RSmola}
     %\end{subfigure}
		%
		%\begin{subfigure}[b]{0.45\textwidth}
         %\centering
		%\includegraphics[width=1\textwidth]{Plots/R_beide_DUmittel.pdf}
		%\caption{}
		%\label{fig:RSmolb}
     %\end{subfigure}
		%\includegraphics[width=0.45\textwidth]{Plots/R.pdf}\\
		%\includegraphics[width=0.45\textwidth]{Plots/R_Kramer.pdf}
		\centering
		\includegraphics[width=0.45\textwidth]{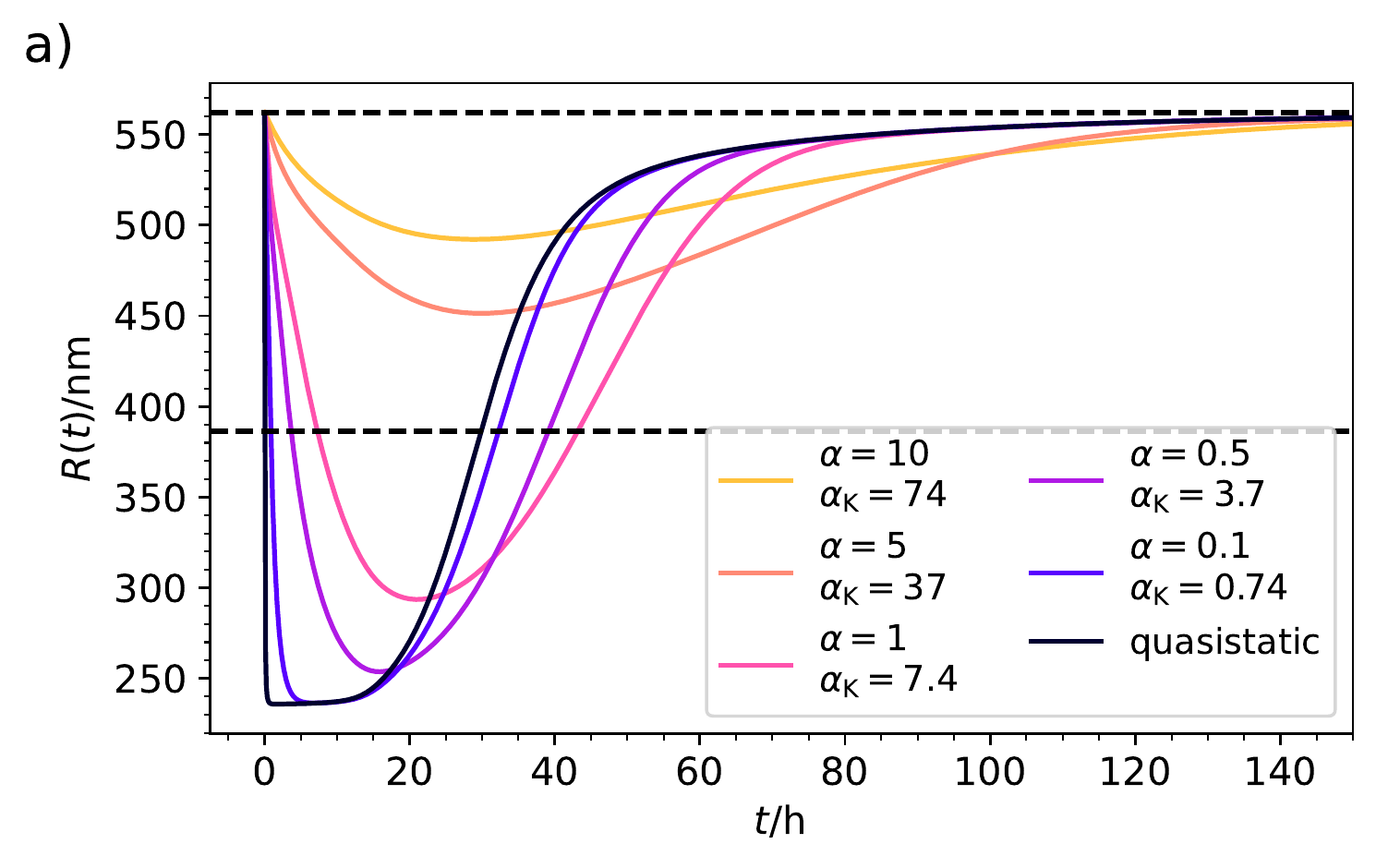}\\
		\includegraphics[width=0.45\textwidth]{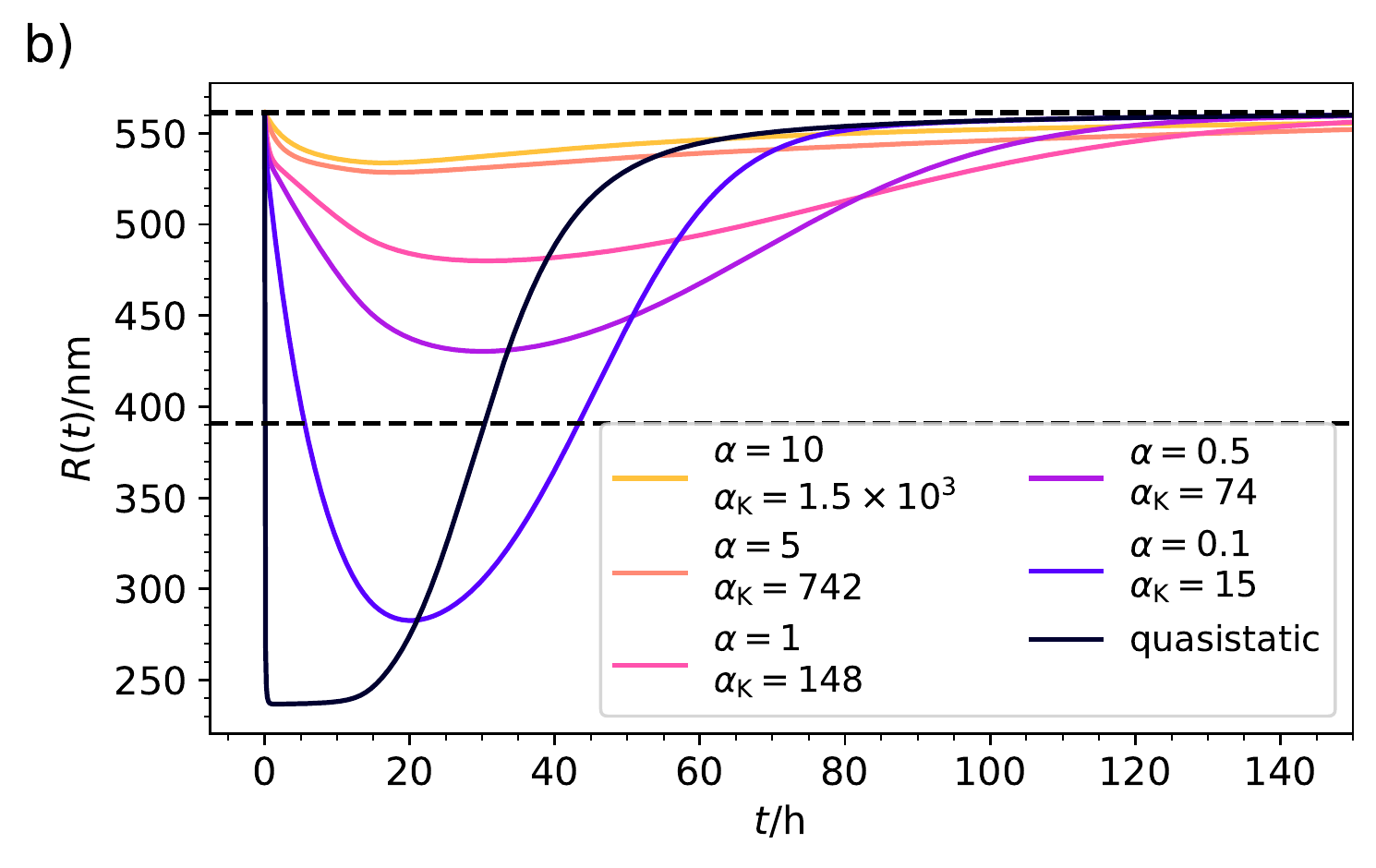}
	\caption{(a) Exemplary results for the time evolution of hydrogel radius $R(t)$ in the case of slow system relaxation calculated from the Smoluchowski approach. {Values for the energy landscape were taken from Table~\ref{tab:fit_results} (for pH$ = 6.1$, $f_0 = 0.99$~mmol/l).} The timescale ratio $\alpha=\tau_\mathrm{D} k_+$ defines the ratio of a typical system diffusion time and the chemical timescale $k_+^{-1}$. {Additionally, we provide the Kramers reference $\alpha_\mathrm{K}$ defined by eq.~(\ref{eq:Kramers}).} For very fast relaxation $\alpha \ll1$ we reach the QS limit. {By the horizontal dashed lines we show the radius before fueling $R\left(t=0 \right)$ and the threshold radius $R(p^*)$.} (b) Same as (a) but now the parameters are chosen such that the transition barrier is {increased from $2\,k_\mathrm{B}T$ to $5\,k_\mathrm{B}T$}.  Here, the system relaxation is too sluggish for the considered $\alpha$ from which we can deduce that $\alpha_\mathrm{K}$ provides the more precise time scale estimate.}
	\label{fig:RSmol}
\end{figure}

In the just described system we have still a relatively small barrier of $\Delta {\cal H} = 2\,k_\mathrm{B}T$. If we raise the barrier, the system relaxation time should be more adequately described by the Kramers crossing time, eq.~(\ref{eq:Kramers}), leading to the time scale ratio $\alpha_\mathrm{K} = \tau_{\rm Kramer}k_+$. This is exemplified in Fig.~\ref {fig:RSmol}(b) where we use the alternative fit with a barrier height of $\Delta {\cal H} = 5\,k_\mathrm{B}T$. The Kramers time is now about 150 times larger than $\tau_\mathrm{D}$ resulting, for example, in $\alpha_\mathrm{K} = 15$ for $\alpha=0.1$ for a response which is already clearly non-QS (brown curve in Fig.~\ref {fig:RSmol}(b)). Hence, the Kramers time is naturally more appropriate to characterize delayed systems involving internal barriers. 

%Since Arrhenius \cite{Arrhenius_1889} we know that the barrier crossing scales exponentially with the barrier. This is the reason, why $\alpha_\mathrm{K}$ is the better choice when discussing the crossing time scale. It should be noted here, that defining $\alpha_\mathrm{K}$ is only possible to a limited extent in our model because the barrier height changes with time. Whenever we calculate values for $\alpha_\mathrm{K}$ we use the case $p=p^*$ meaning that we neglect the impact of the $m$ value in this way. This might be missleading especially for large $m$ values. Here the difference between one minimum and the maximum at $p_\mathrm{max}$ can be much smaller than in the symmetric case leading to a transition faster than estimated by our $\tau_\mathrm{K}$. In our energylandscape obtained by quasistatic fits, only the difference between the minima is relevant so that $\left<R\right>$ is weakly affected by the barrier height between them. This means, that an increased $\Delta U$ is within our fits not compensated by $m$, so that the barrier is indeed larger, so that $\alpha_\mathrm{K}$ is in general a good estimate for the crossing timescale.}
%This indicates, that the time needed for barrier crossing exceeds the timescales observed in the experiment.}

\section{Concluding remarks}

In this contribution, we have put forward a Landau type of mean-field model to describe the quasi-static and nonequilibrium chemical fueling of transient soft matter states in a bistable system.  Already the analysis of the quasi-static (QS) limit led to useful scaling laws and relations between chemical and mechanical parameters, which could serve for future material design. We demonstrated their usefulness explicitly for the case study of the chemically fueled volume (collapse) transition of a responsive hydrogel colloid. Moreover, we provided a thermodynamic (energy, work, and power) analysis in the QS limit and also showed how internal (diffusive) relaxation times scales can substantially alter the time evolution if they compete with the chemical timescales of fueling.  

Several extensions of this model shall be interesting in future studies.  By including more complex order parameters, higher dimensions of the reaction coordinate, or structural gradients in the Hamiltonian, the extension to self-assembling~\cite{heuser} and phase separating systems~\cite{deng,heckel} could be attempted. From the chemical side, higher order reaction networks \cite{Hoefling,Deng_2021,Klemm_2022} than only two-step reactions could be envisioned. Finally,  further increasing complexity could be obtained by imposing a negative feedback cycle within the system, e.g., by coupling the mechanical response back to the chemical reaction~\cite{Li2000,Bell2021a, Jain_2021}. Here, much more intricate transient dynamics and responses, including regimes of mono- and bistability, excitability, damped oscillations, as well as sustained oscillatory states can be expected during the time evolution~\cite{Hoefling,Abeer}.

% -------------------------------------------------------
% -------------------------------------------------------
	\section{Acknowledgments}

The authors thank Andreas Walther for useful discussions and sharing details of the fueling experiments on hydrogels. The authors also thank Nils G\"oth and Sebastian Milster for a critical reading of the manuscript. The authors further acknowledge support by the state of Baden-Württemberg through bwHPC and the German Research Foundation (DFG) through grant no INST 39/963-1 FUGG (bwForCluster NEMO) and  funding from the DFG under Germany's Excellence Strategy - EXC-2193/1 - 390951807 (`LivMatS').

% -------------------------------------------------------
% -------------------------------------------------------
%\appendix
%\footnote{The equilibration time is extended to $5\cdot10^6$ timesteps for systems with both a high density ($\rho\sigma_0^3 = 0.95, 1.33$) and a low switching rate ($\lambda\tau = 0.01, 0.025$).}

% -------------------------------------------------------

	% The \nocite command causes all entries in a bibliography to be printed out
	% whether or not they are actually referenced in the text. This is appropriate
	% for the sample file to show the different styles of references, but authors
	% most likely will not want to use it.
	%\nocite{*}
	
	\bibliographystyle{apsrev4-2-no-url.bst}

\end{document}